\documentclass[aps,pra,twocolumn,amsmath,amssymb,superscriptaddress,linenumbers]{revtex4-1}

\usepackage{graphicx}
\usepackage{bm}
\usepackage{dcolumn}
\usepackage{amsmath}
\usepackage{verbatim}     
\usepackage[dvipsnames]{xcolor}       
\usepackage[normalem]{ulem}   

\newcommand{\beq}{\begin{equation}}
\newcommand{\eeq}{\end{equation}}
\newcommand{\bei}{\begin{itemize}}			
\newcommand{\eei}{\end{itemize}}			

\newcommand{\hg}[1]{\hbox{HyGG}_{#1}}		

\usepackage{hyperref}
\hypersetup{%
   pdfpagemode=None, 
   pdfstartpage=1,
   pdfmenubar=true,
   pdftoolbar=true,
   colorlinks = true,
   linkcolor=blue,
   citecolor=blue,
   urlcolor=blue,
   bookmarksopen=false
 }

\begin{document}

\nolinenumbers
\title{Measuring the complex orbital angular momentum spectrum and spatial mode decomposition of structured light beams}
\author{Alessio D'Errico}
\affiliation{Dipartimento di Fisica, Universit\`{a} di Napoli Federico II, Complesso Universitario di Monte Sant'Angelo, Via Cintia, 80126 Napoli, Italy}
\author{Raffaele D'Amelio}
\affiliation{Dipartimento di Fisica, Universit\`{a} di Napoli Federico II, Complesso Universitario di Monte Sant'Angelo, Via Cintia, 80126 Napoli, Italy}
\author{Bruno Piccirillo}
\affiliation{Dipartimento di Fisica, Universit\`{a} di Napoli Federico II, Complesso Universitario di Monte Sant'Angelo, Via Cintia, 80126 Napoli, Italy}
\author{Filippo Cardano}\email{filippo.cardano2@unina.it}
\affiliation{Dipartimento di Fisica, Universit\`{a} di Napoli Federico II, Complesso Universitario di Monte Sant'Angelo, Via Cintia, 80126 Napoli, Italy}
\author{Lorenzo Marrucci}
\affiliation{Dipartimento di Fisica, Universit\`{a} di Napoli Federico II, Complesso Universitario di Monte Sant'Angelo, Via Cintia, 80126 Napoli, Italy}
\affiliation{CNR-ISASI, Institute of Applied Science and Intelligent Systems, Via Campi Flegrei 34, Pozzuoli (NA), Italy}

\begin{abstract} 
Light beams carrying orbital angular momentum are key resources in modern photonics. In many applications, the ability of measuring the complex spectrum of structured light beams in terms of these fundamental modes is crucial. Here we propose and experimentally validate a simple method that achieves this goal by digital analysis of the interference pattern formed by the light beam and a reference field. Our approach allows one to characterize the beam radial distribution also, hence retrieving the entire information contained in the optical field. Setup simplicity and reduced number of measurements could make this approach practical and convenient for the characterization of structured light fields.

\end{abstract}

\maketitle


\noindent In 1992 Allen et al. \cite{Allen1992} showed that helical modes of light -- paraxial beams featuring an helical phase factor $e^{i m \phi}$, where $\phi$ is the azimuthal angle around the beam axis and $m$ is an integer -- carry a definite amount of orbital angular momentum (OAM) along the propagation axis, equal to $m\hbar$ per photon \cite{Yao2011}. Important realizations of such optical modes are, for example, Laguerre-Gauss (LG) \cite{Allen1992} and Hypergeometric Gaussian beams (HyGG) \cite{Karimi2007}, which share the typical twisted wavefront but differ in their radial profiles. Controlled superpositions of helical modes, possibly combined with orthogonal polarization states via spin-orbit interaction~\cite{Bliokh2015,Cardano2015}, result in spatially structured beams that are proving useful for a broad set of photonic applications~\cite{Rubinsztein-Dunlop2017}, such as for instance classical and quantum optical communication~\cite{Vallone2014,Willner2015,Mirhosseini2015,Bouchard2017}, quantum information processing~\cite{Wang2015,Malik2016,Hiesmayr2016}, and quantum simulations \cite{Cardano2016,Cardano2017}.
The ability to ascertain experimentally the OAM values associated with individual helical modes represents a fundamental requirement for all applications based on twisted light. Hitherto, this has been demonstrated by a variety of methods: exploiting double slit interference \cite{Sztul2006}, diffraction through single apertures \cite{Hickmann2010,Ferreira2011,Mourka2011,Mazilu2012} or through arrays of pinholes \cite{Berkhout2008}, interference with a reference wave \cite{Harris1994,Padgett1996}, interferometers \cite{Leach2002,Slussarenko2010,Lavery2011}, OAM-dependent Doppler frequency shifts \cite{Courtial1998,Vasnetsov2003,Zhou2016}, phase flattening and spatial mode projection using pitchfork holograms \cite{Mair2001,Kaiser2009,Schulze2013}, q-plates \cite{Karimi2009a,Karimi2012}, spiral phase plates \cite{Bierdz2013} and volume holograms \cite{Gruneisen2011}, spatial sorting of helical modes by mapping OAM states into transverse momentum (i.e. propagation direction)
 \cite{Berkhout2010,Mirhosseini2013}, quantum weak measurements \cite{Malik2014}. 
 
 \noindent General structured fields are however not given by individual helical modes, but can always be obtained as suitable superpositions of multiple helical modes. Accordingly, a full experimental characterization of these structured fields can be based on measuring the complex coefficients (amplitude and phase) associated with each mode appearing in the superposition, for any given choice of the mode basis. In general, this is not a trivial task, but several methods for the reconstruction of the complex spectrum associated with the OAM degree of freedom have been demonstrated thus far \cite{Leach2002,Karimi2009a,Bierdz2013,Gruneisen2011,Vasnetsov2003,Zhou2016,Forbes2016,Zhao2017}, possibly including also the radial mode spectrum reconstruction \cite{Mirhosseini2013,Berkhout2010,Malik2014,Schulze2013,Kaiser2009,Berkhout2011,Mirhosseini2016,Fickler2017}. It is worth noting that, once these complex coefficients are known, the complete spatial distribution of the electric field can be obtained and important properties such as  beam quality factor $M^2$, beam width and wavefront are easily computed at any propagation distance \cite{Schulze2012,Forbes2016}. Inspired by previous works \cite{Abouraddy2011,Vasnetsov2003,Zhou2016} introducing Fourier analysis in this context, here we present an approach to the measurement of light OAM spectrum and, more generally, to spatial mode decomposition of structured light that may prove to be more practical than most alternatives. The OAM complex spectrum information is contained in the intensity pattern resulting from the interference of the light beam with a known reference field (such as a Gaussian beam), and can be hence easily extracted by a suitable processing of the corresponding images recorded on a camera. First, Fourier transform with respect to the azimuthal angle leads to determining the complex coefficients associated with each OAM value, as a function of the radial coordinate. Numerical integration over the latter then allows one to use this information to determine the OAM power spectrum and, eventually, to decompose each OAM component in terms of radial modes, e.g. LG beams. Remarkably, the whole information associated with the spatial mode decomposition, or with the OAM power spectrum, is contained in a few images, whose number does not scale with the dimensionality of the set of detected helical modes. A unique series of data recorded for the characterization of a given field is used for obtaining the decomposition in any basis of spatial modes carrying OAM (LG, HyGG, Bessel,...), as this choice comes into play only at the stage of image analysis.

\section*{Results and Discussion}
\noindent {\bf Description of the technique.} In the following, we limit our attention to the case of scalar optics, as the extension to the full vector field is simply obtained by applying the same analysis to two orthogonal polarization components. Considering cylindrical coordinates $(r,\phi,z)$, the electric field amplitude associated with a monochromatic paraxial beam propagating along the $z$ direction is given by:
\begin{align}\label{eq:paraxialbeam}
E_s(r,\phi,z,t)=A_s(r,\phi,z)e^{-i(\omega t-k\,z) },
\end{align}
where $\omega$ is the optical frequency and $k$ is the wave number. We refer to $E_s$ as the signal field, to distinguish it from the reference beam that will be introduced later on. The information concerning the spatial distribution of the field is contained in the complex envelope $A_s(r,\phi,z)$. Being periodic with respect to the azimuthal coordinate $\phi$, such complex function can be expanded into a sum of fundamental helical modes $e^{i m \phi}$, carrying $m\hbar$ OAM per photon along the $z$ axis \cite{Yao2011}:
\begin{align}\label{eq:helicaldecomposition}
A_s(r,\phi,z)=\sum_{m=-\infty}^{\infty}c_m(r,z)e^{im\phi},
\end{align}
where coefficients $c_m$ are defined in terms of the angular Fourier transform
\begin{align}\label{eq:coefficients}
c_m(r,z)=\frac{1}{2\pi}\int_0^{2\pi} d\phi\, e^{-im\phi}A_s(r,\phi,z).
\end{align}
The probability $P(m)$ that a photon is found in the $m$-order OAM state is obtained from the coefficients $c_m$ by integrating their squared modulus along the radial coordinate:
\begin{align}\label{eq:OAMprobability}
P(m)=\frac{1}{S}\int_{0}^{\infty} d r\,r\,|c_m(r,z)|^2,
\end{align}
where $S=\sum_m\int_{0}^{\infty} d r\,r\,|c_m(r,z)|^2$ is the beam power at any transverse plane. The quantity $P(m)$ is also referred to as the OAM power spectrum, or spiral spectrum of the beam, and does not depend on the longitudinal coordinate $z$, because of OAM conservation during propagation. A complete analysis of the field in terms of transverse spatial modes is obtained by replacing $e^{im\phi}$ in Eq.\ \ref{eq:helicaldecomposition} with a complete set of modes having a well defined radial dependance, e.g. LG modes:
\begin{align}\label{eq:spatialdecomposition}
A(r,\phi,z)=\sum_{p=0}^{\infty}\sum_{m=-\infty}^{\infty}b_{p,m}\, \text{LG}_{p,m}(r,\phi,z),
\end{align}
where $\text{LG}_{p,m}(r,\phi,z)$ is the complete LG mode of integer indices $p$ and $m$ (with $p\geq0$), as explicitly defined in the Methods. The link between coefficients $c_m$ and $b_{p,m}$ is then given by:
\begin{align}\label{eq:coefficients}
b_{p,m}=\int_0^\infty r\,dr\,\text{LG}_{p,m}^*(r,z)\,c_m(r,z),
\end{align}
where we introduced the radial LG amplitudes $\text{LG}_{p,m}(r,z)=\text{LG}_{p,m}(r,\phi,z) e^{-i m\phi}$, for which the $\phi$ dependence is removed.

The procedure we present here allows one to measure the complex quantities $c_m(r)$, or equivalently the coefficients $b_{p,m}$. We achieve this goal by letting the signal optical field interfere with a reference wave $E_{\text{ref}}=A_{\text{ref}}(r,\phi,z)e^{-i(\omega t-kz)}$, having the same polarization, frequency, wavelength and optical axis of the beam under investigation, and whose spatial distribution is known. The simplest choice for this reference is a Gaussian beam. At any plane transverse to the propagation direction, the intensity pattern $I$ formed by the superposition of signal and reference beams is (we omit the functional dependance on the spatial coordinates)
\begin{align}\label{eq:intensitypattern}
I=&I_s+I_{\text{ref}}+\widetilde I_\alpha.
\end{align}
Here $I_s$ and $I_\text{ref}$ are the intensities corresponding to the sole signal and reference fields, respectively, while the term $\widetilde I_{\alpha}=2\,\text{Re}{(e^{i\alpha}A_s\,A_\text{ref}^*)}$ corresponds to their interference modulation pattern, $\alpha$ being a controllable optical phase between the two. The interference modulation pattern can be experimentally singled out by taking three images, namely $I$, $I_{\text{ref}}$ (blocking the signal beam) and $I_s$ (blocking the reference beam), and then calculating the difference $\widetilde I_{\alpha}=I-I_{\text{ref}}-I_s$.

The interference modulation pattern is linked to the OAM mode decomposition by the following expression:
\begin{align}\label{eq:interference}
\widetilde I_\alpha=2\sum_{m} |A_{\text{ref}}||c_m|\,\cos{[m\phi+\alpha+\beta_m]},
\end{align}
where $\beta_m(r,z)=\text{Arg}[c_m(r,z)]-\text{Arg}[A_{\text{ref}}(r,z)]$. By combining two interference patterns obtained with $\alpha=0$ and $\alpha=\pi/2$ one then gets:
\begin{align}\label{eq:interferencefinal}
\widetilde I_0-i\, \widetilde I_{\pi/2}=2\sum_{m}|A_{\text{ref}}||c_m|\,e^{i[m\phi+\beta_m]}
\end{align}
Finally, Fourier analysis with respect to the azimuthal coordinate allows one to determine the coefficients $c_m(r)$:
\begin{align}\label{eq:interferencefinalcoefficients}
c_{m}(r,z)=\frac{1}{4\pi A_{\text{ref}}^*(r,z)}\int_0^{2\pi}d\phi(\widetilde I_0-i\,\widetilde I_{\pi/2})e^{-im\phi},
\end{align}
which contains all the information associated with the spatial distribution of the electric field.

The method just described is required for a full modal decomposition and requires taking a total of four images (that is $I$ with $\alpha=0$ and $\alpha=\pi/2$, plus $I_{\text{ref}}$ and $I_s$), maintaining also a good interferometric stability between them. However, for applications requiring the measurement of the OAM power spectrum only, that is ignoring the radial structure of the field, and for which the OAM spectrum is bound from below (that is, there is a minimum OAM value) there is a simplified procedure that is even easier and more robust (the case for which the spectrum is limited from above can be treated equivalently). This usually requires to have the signal beam pass first through a spiral optical phase element, described by the transfer factor $e^{iM\phi}$ (this can be achieved with a $q$-plate or a spiral phase plate with the appropriate topological charge). If $M$ is higher than the absolute value of the minimum negative OAM component of the signal beam, the spiral spectrum of the beam after this optical component will contain only modes associated with positive OAM values. Then, one can extract the associated probabilities $P(m)$ by Fourier analysis of $\widetilde I_0$ only (see Eq.\ \ref{eq:interference}), with no need of measuring also $\widetilde I_{\pi/2}$, thus reducing the number of required images to three. We discuss this in detail in the final part of the paper, showing also that such procedure is intrinsically less sensitive to noise corresponding to beam imperfections.

\begin{figure}[h!]
\fbox{\includegraphics[width=\linewidth]{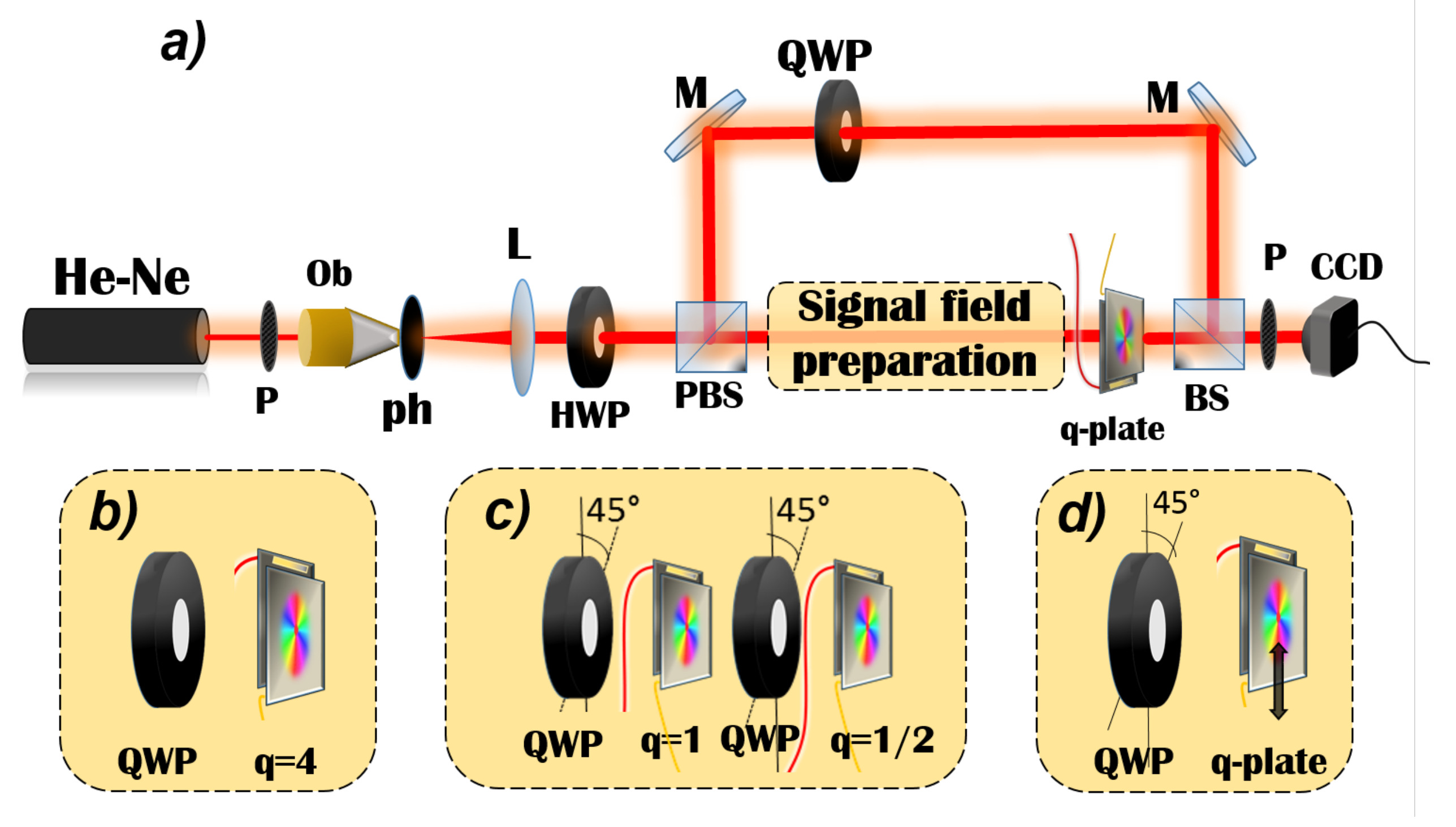}}
\caption{Sketch of the experimental apparatus. {\bf a)} A He-Ne laser beam passes through a polarizer (P) and is spatially cleaned and collimated by means of an objective (Ob), a pinhole (ph) and a lens (L). A half-wave plate (HWP) and a polarizing beam splitter (PBS) are used in order to split the beam into the signal and reference arms, whose relative intensities can be controlled by HWP rotation. Fields resulting from a complex superposition of multiple helical modes were obtained by using $q$-plates and quarter-wave plates (QWPs), as shown in panels b-c. After preparing the signal field, we place a further sequence of a QWP and a $q$-plate in case we need to shift the entire OAM spectrum. The reference field is a $\text{TEM}_{0,0}$ Gaussian mode. In the upper arm of the interferometer, by orienting the QWP at 0 or $90^\circ$ with respect to the beam polarization we can introduce a $\alpha=0$ or $\pi/2$ phase delay between the signal and the reference field, respectively. The two beams are superimposed at the exit of a beam splitter (BS) and filtered through a polarizer, so that they share the same polarization state. The emerging intensity pattern is recorded on a CCD camera (with resolution $576\times668$). \textbf{b)} A QWP oriented at $45^\circ$ or 0, followed by $q$-plate with $q=4$ and $\delta=\pi$ or $\delta=\pi/2$, is used for the generation of a light beam containing a single mode ($m=8$) or three modes $(m=-8,0,8)$, respectively. {\bf c)} two $q$-plates with $q=1$ and $q=1/2$ are aligned to generate spectra with $m\in[-3,3]$. {\bf d)} A set of more complex distributions was obtained by displacing laterally the centre of a $q$-plate ($q=1$ and $\delta=\pi$) with respect to the axis of the impinging Gaussian beam.}
\label{fig: exp_setup}
\end{figure}
\begin{figure}[t!]
\centering
\fbox{\includegraphics[width=\linewidth]{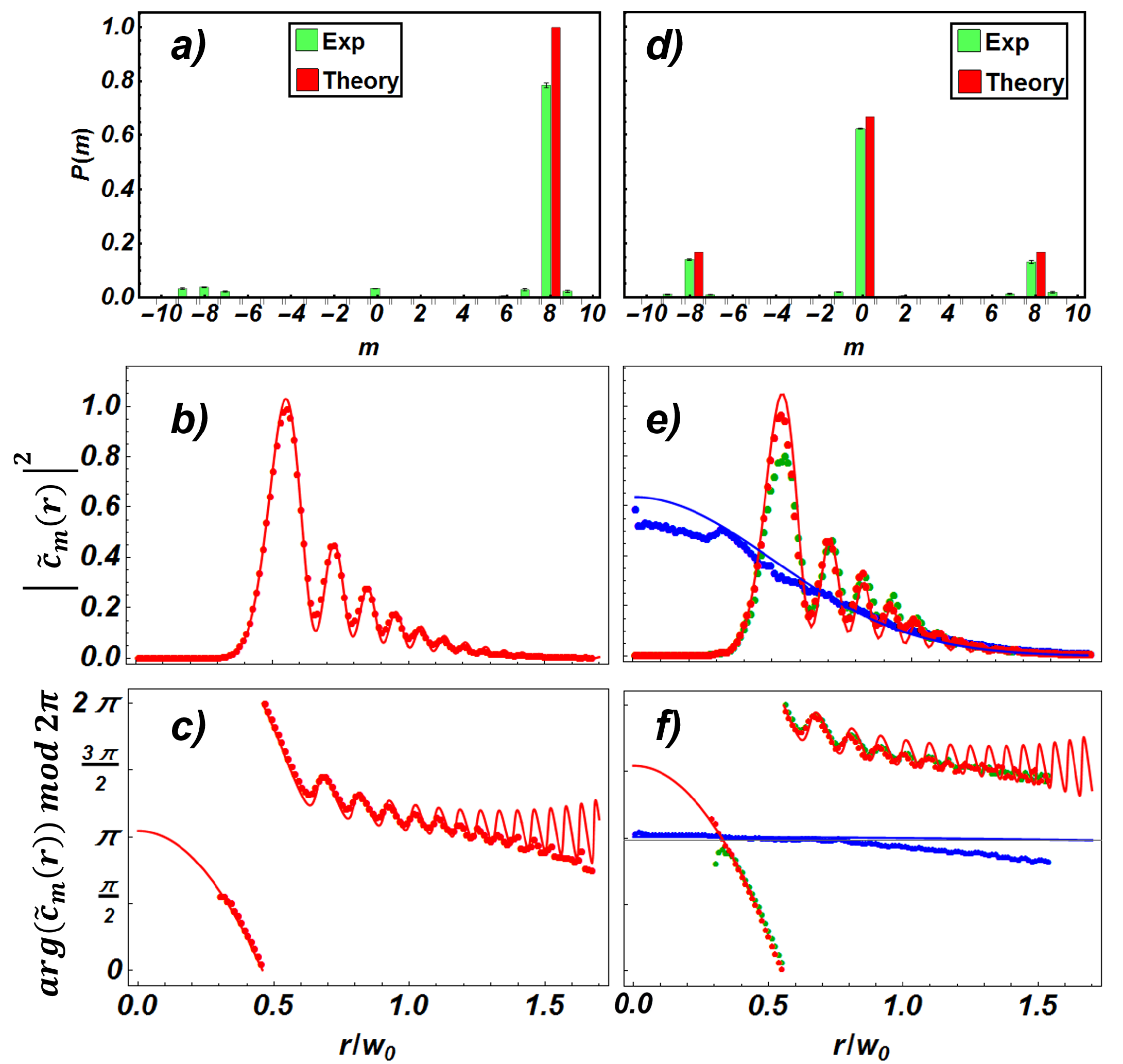}}
\caption{Experimental reconstruction of light OAM spectrum. 
We report the experimental characterization of optical fields containing one (a-b-c) and three (d-e-f) helical modes, generated using a $q$-plate with $q=4$ and $\delta=\pi$ or $\pi/2$, respectively. Panels a and d show the OAM distributions in the two cases. Error bars are calculated as three times the standard error. Panels b,c and e,f show the measured amplitude and phase profiles of the non-vanishing helical modes that are present in the beam, where blu, red and green coloured points are associated with modes with $m=0,8,-8$, respectively. These results are compared with theoretical simulations, represented as continuous curves with the same color scheme adopted for the experimental results. For each value of $m$, we plot normalized coefficients $\tilde c_m=c_m/S_m$, where $S_m$ is the total power associated with the helical mode. As expected from theory, a fraction of the beam is left in the fundamental Gaussian state, while an equal amount of light is converted into helical modes with $m=\pm 8$, both having the radial profile of a HyGG$_{-8,8}$ mode. Simulated profiles of Gaussian and HyGG modes correspond to $w_0=1.45$ mm and $z=30$ cm, the latter being the distance between the $q$-plate and the camera. Error bars are smaller than experimental points. 
}
\label{fig: amp_phase_h_d=p2.pdf}
\end{figure}
\noindent {\bf Experimental results.} We demonstrate the validity of our technique by determining the OAM spectrum and the radial profile of the associated helical modes for a set of structured light fields. The setup is shown in Fig.\ref{fig: exp_setup} and described in detail in the figure caption. Here, structured light containing multiple OAM components is generated by means of $q$-plates, consisting essentially in a thin layer of liquid crystals whose local optic axes are arranged in a singular pattern, characterized by a topological charge $q$ \cite{Marrucci2006,Marrucci2011}. The way such device modifies the spatial properties of a light beam is described in detail in the Methods section.

\noindent In Figs.\ \ref{fig: amp_phase_h_d=p2.pdf}a-c and \ref{fig: amp_phase_h_d=p2.pdf}d-f we report the results of our first experiment, consisting in the measurement of both amplitude and phase of coefficients $c_m(r)$ of optical fields having one ($m=8$) and three ($m=\{-8,0,8\}$) different helical modes, respectively, accompanied by the associated OAM power spectrum (see Eq.\ \ref{eq:OAMprobability}). We generate such structured light by shining a $q$-plate with $q=4$ with a left-circularly (horizontally) polarized Gaussian beam and setting the plate optical retardation $\delta$ to the value $\pi$ ($\pi/2$), respectively (see the Methods). Our data nicely follow the results from our simulations, with some minor deviations that are due to imperfections in the preparation of the structured fields. In particular, in panel a, the small peaks centered around $m=-8$ are related to the possible ellipticity of the polarization of the Gaussian beam impinging on the $q$-plate, while a small contribution at $m=0$ corresponds to the tiny fraction of the input beam that has not been converted by the $q$-plate. In panels b and e the radial profiles used for our simulations are those corresponding to the Hypergeometric-Gaussian modes \cite{Karimi2007}, the helical modes that are expected to describe the optical field at the exit of the $q$-plate \cite{Karimi2009} (see the Methods for details). Error bars shown in our plots are those associated with the variability in selecting the correct center $r=0$ in the experimental images, which is identified as one of the main source of uncertainties in the spectral results. They are estimated as three times the standard deviation of the data computed after repeating our analysis with the coordinate origin set in one of 25 pixels that surround our optimal choice. Other possible systematic errors, such as for example slight misalignments between the signal and reference fields, are not considered here.
 
\begin{figure}[t!]
\centering
\fbox{\includegraphics[width=\linewidth]{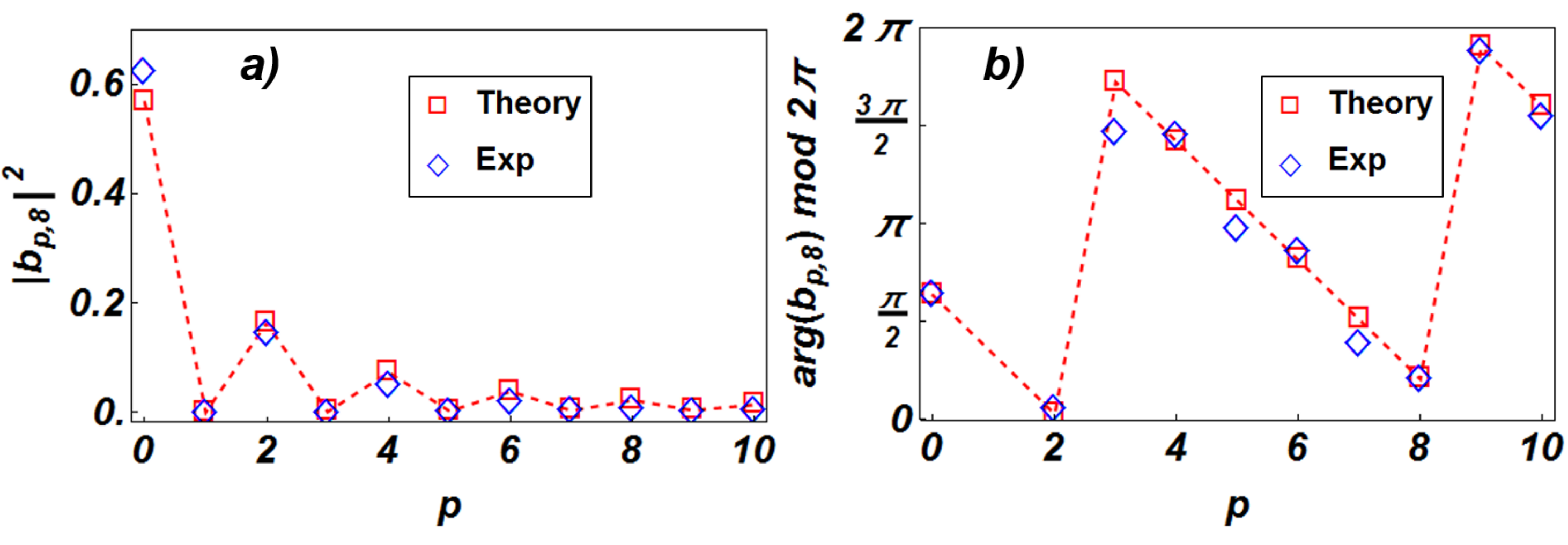}}
\caption{Complete spatial mode decomposition in terms of LG beams. We consider the light beam emerging from a $q$-plate with ($q=4$, $\delta=\pi$), described by a HyGG$_{-8,8}$ mode \cite{Karimi2009}. We evaluate the overlap integral between the radial envelope $c_8(r)$ measured in our experiment at $z=30$ cm and LG$_{p,8}$ modes at the same value of $z$ and characterized by the optimal beam waist $\widetilde{w}_0=w_0/9$ \cite{Vallone2017}, where $w_0$ is the input beam waist. In panels {\bf a)} and {\bf b)} we plot the squared modulus and the phase of the resulting coefficients (blue markers), respectively, showing a good agreement with the values obtained from numerical simulations (red markers). The phase corresponding to $b_{1,8}$ is absent in the plot since the corresponding coefficient amplitude is vanishing.
}
\label{fig:LGexpansion}
\end{figure}

\begin{figure}[b!]
\centering
\includegraphics[width=\linewidth]{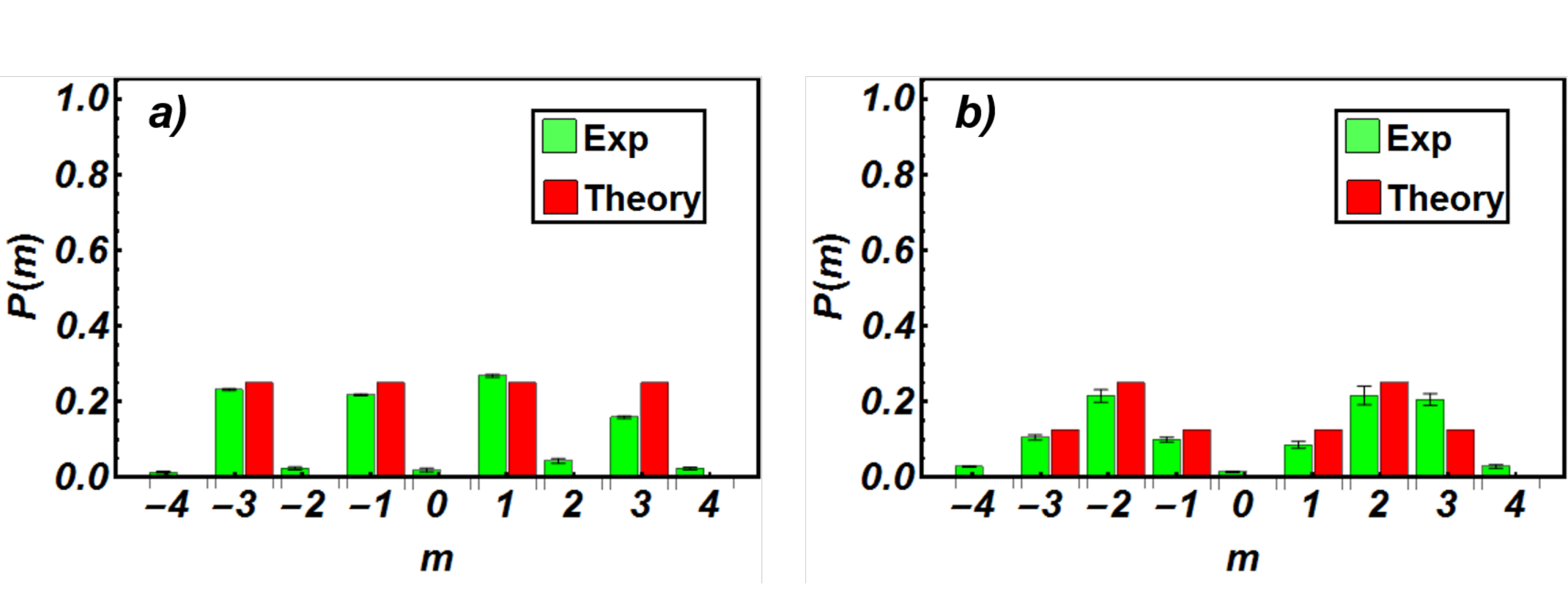}
\caption{Measure of shifted OAM power spectrum. OAM probability distributions are measured for two different optical fields, obtained when shining a sequence of two $q$-plates with $q_1=1$ and $q_2=1/2$. A further $q$-plate with $q=4$ shifts the final spectrum by $M=8$ units. {\bf a)} OAM spectrum for the case $\delta_1=\pi$ and $\delta_2=\pi$. {\bf b)} The same data are reported for a different field, obtained when $\delta_1=\pi$ and $\delta_2=\pi/2$. Error bars represent the standard error multiplied by three.}
\label{fig: 3q-plate}
\end{figure}

\begin{figure}[]
\centering
\includegraphics[width=\linewidth]{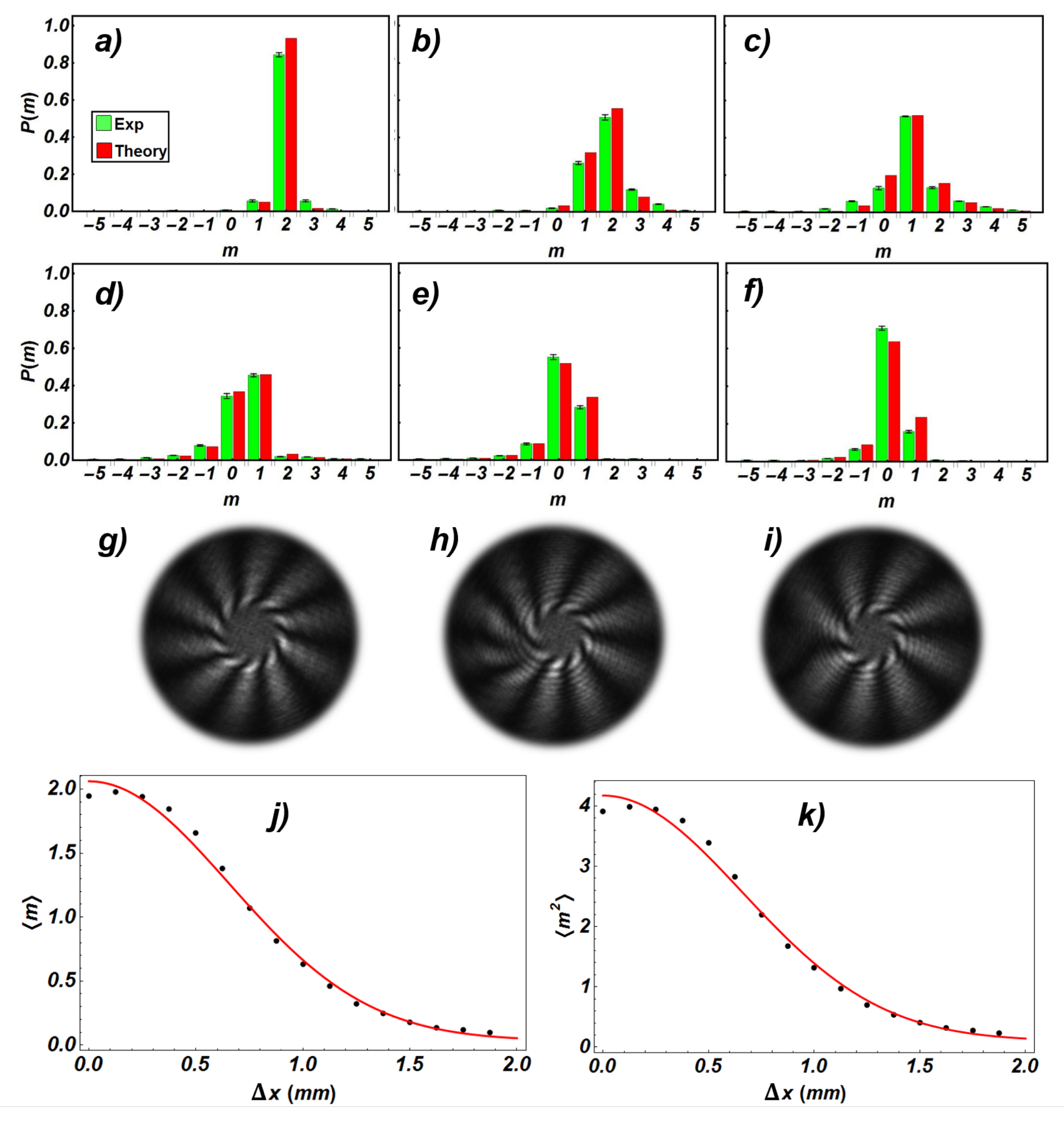}
\caption{OAM spectrum for a shifted $q$-plate. We measure the OAM power spectrum at the exit of a $q$-plate ($q=1$, $\delta=\pi$) shifted with respect to the axis of the impinging Gaussian beam, which is left-circularly polarized. The overall spectrum is shifted by $M=8$ units since we are using a further $q$-plate with $q=4$ and $\delta=\pi$. However, we plot the original OAM distribution associated with the signal field. {\bf a-f)} Experimental (green) and simulated (red) OAM power spectra when the lateral shift is equal to $a\Delta x$, with $a=1,3,6,9,12,15$ and $\Delta x$ =0,125 mm, respectively. Error bars represent the standard error multiplied by three. {\bf g-i)} Examples of the experimental intensity pattern $I_0$ used for determining the power spectra reported in panels a,c,e. The number of azimuthal fringes reveals that the OAM spectrum has been shifted. {\bf j-~k)} First and second moment ($\langle m \rangle$ and $ \langle m^2 \rangle$) measured as a function of the lateral displacement. Error bars are not visible because smaller than the experimental points.} 
\label{fig: q-platedisp}
\end{figure}

\begin{figure}[t]
\centering
\includegraphics[width=8 cm]{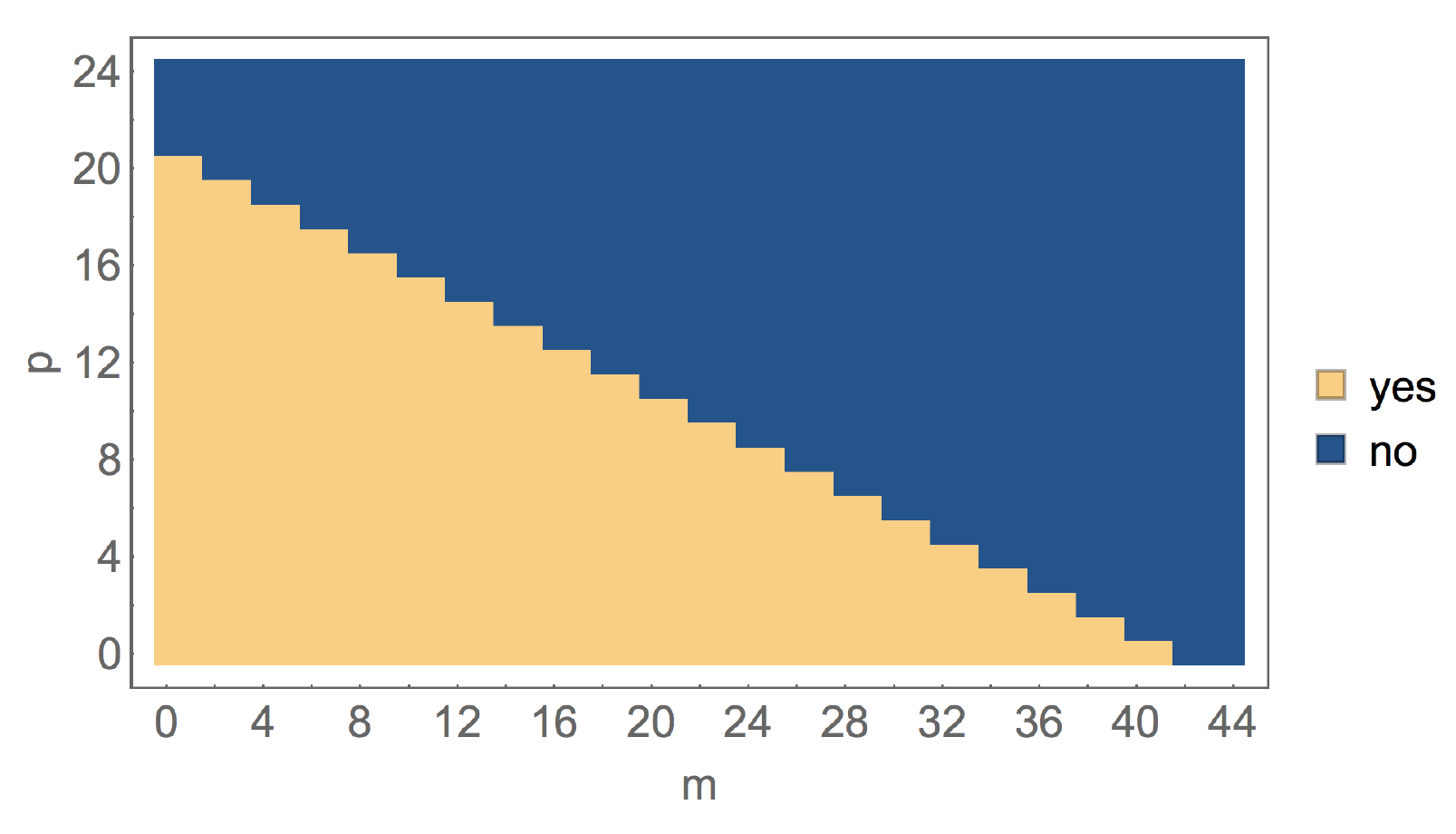}
\caption{Detectable LG modes. Different colors, as reported in the legend, indicate whether a specific LG$_{p,m}$, in a transverse plane $z=30$ cm and with a beam waist $w_0=0.16$ mm [$w(z=30$ cm)$\rightarrow 0.4$ mm] can be resolved in our setup. These parameters correspond to the ones used for the complete spatial decomposition of the HyGG beam generated by a $q$-plate ($q=4,\delta=\pi$) in terms of LG beams.}
\label{fig:LG_detectable}
\end{figure}

\noindent Data reported in Fig.\ \ref{fig: amp_phase_h_d=p2.pdf} prove our ability to measure the complex radial distribution of the field associated with individual helical modes in a superposition. For each of these, we can use our results to obtain a decomposition in terms of a complete set of modes. For a demonstration of this concept, we consider the field obtained when a left-circularly-polarized beam passes through a $q$-plate with $q=4$ and $\delta=\pi$. The latter contains only a mode with $m=8$, as shown in Fig.\ \ref{fig: amp_phase_h_d=p2.pdf}a-c. By evaluating the integrals reported in Eq.\ \ref{eq:coefficients} we determine the coefficients $b_{p,8}$ of a LG decomposition. For our analysis, we use LG beams with an optimal waist parameter $\widetilde w_0$ (different from the one of the impinging Gaussian beam), defined so that the probability of the lowest radial index $p=0$ is maximal \cite{Vallone2017}. In Fig.\ \ref{fig:LGexpansion} we plot squared modulus and phase of the coefficients $b_{p,8}$ determined experimentally, matching nicely the results obtained from numerical simulations.\\[1 ex]
{\bf Shifting the OAM power spectrum.} As mentioned above, shifting the OAM spectrum of the signal field may be used to simplify its measurement, when reconstructing the radial profile is not needed. In our case, we let the signal field pass through a $q$-plate with $q=M/2$ and $\delta=\pi$, after preparing it in a state of left-circular polarization. If $M$ is large enough, i.e. higher than the smallest OAM component of the signal field, we have that $c(m)\neq0$ only if $m>0$. This allows in turn using Eq.\ \ref{eq:interference} to determine the OAM spectrum, instead of Eq.\ \ref{eq:interferencefinal} that requires the measurement of $I_{\pi/2}$ also. At the same time, this approach is less sensitive to possible noise due to beam imperfections, typically associated with small spatial frequencies and affecting lowest-order helical modes, as reported also in Refs.\ \cite{Vasnetsov2003,Zhou2016}. Let us note that once the beam passes through an optical element adding the azimuthal phase $e^{iM\phi}$, thanks to the conservation of OAM during free propagation, the associated power OAM spectrum is only shifted by $M$ units, that is $P(m)\rightarrow P(m+M)$. The radial distribution of individual helical modes, on the other hand, is altered during propagation, that is $c_m(r,z)\not \rightarrow c_{m+M}(r,z)$. For this reason, this alternative procedure proves convenient only when determining the OAM probability distribution but cannot be applied to the reconstruction of the full modal decomposition. In Fig.\ \ref{fig: 3q-plate}, we report the measured power spectrum of different fields containing helical modes with $m\in[-3,3]$ (see the Methods and the figure caption for further details on the generation of such complex fields), as determined by shifting the OAM spectrum by $M=8$ by means of a $q$-plate with $q=4$ and $\delta=\pi$.\\[1 ex]
{\bf OAM spectra for displaced $q$-plates.} As a final test, we used our technique for characterizing more complex optical fields, such as those emerging from a $q$-plate whose central singularity is displaced with respect to the input Gaussian beam axis (Fig.\ref{fig: exp_setup}c). In Fig.\ \ref{fig: q-platedisp} we report the OAM probability distributions obtained when translating a $q$-plate ($q=1$, $\delta=\pi$) in a direction that is parallel to the optical table, with steps of $\Delta x=0.125$ mm. Our data are in excellent agreement with results obtained from numerical simulations.  In the same figure we show part of the associated total intensity patterns $I_0$ (see Eq.\ \ref{eq:intensitypattern}) recorded on the camera. In addition, for each configuration we show in Fig.\ \ref{fig: q-platedisp} that  the first and second order moments of the probability distributions are characterized by Gaussian profiles $\langle m \rangle=2q \exp(-2x_0^2/w_0^2)$ and $\langle m^2 \rangle=(2q)^2 \exp(-2x_0^2/w_0^2)$ \cite{Piccirillo2015}. Fitting our data so that they follow the expected Gaussian distributions (red curves) we obtain $w_0^{\text{fit}}=1.36\pm 0.04$ mm from $\langle m \rangle$ (panel g), and $w_0^{\text{fit}}=1.39\pm 0.06$ mm from $ \langle m^2 \rangle$ (panel h), which are close to the expected value $w_0=1.45\pm 0.18$ mm.\\[1 ex]
\textbf{Range of detectable helical modes.} Finite size of the detector area and the camera resolution impose natural limitations to our approach, that cannot be used to characterize helical modes with arbitrary values of $m$ and radial profiles. Starting from these considerations, in the Methods we describe how to evaluate the bandwidth of detectable LG modes in terms of sensor area and resolution, and provide an explicit example for our specific configuration (associated data are reported in Fig.\ \ref{fig:LG_detectable}).
\section*{Conclusions}
\noindent In this study we introduced a new technique for measuring the orbital angular momentum spectrum of a laser beam, accompanied by its complete spatial mode decomposition in terms of an arbitrary set of modes that carry a definite amount of OAM, such as LG beams or others. Based on the azimuthal Fourier analysis of the interference pattern formed by the signal and the reference field, relying on only a few measurements this approach allows one to readily extract the information contained in both the radial and azimuthal degrees of freedom of a structured light beam. The most general method requires taking four images, including the intensity patterns of the signal beam, the reference beam and two interference patterns between them. Information on the modal decomposition of the signal field is then retrieved using a simple dedicated software. Since the spatial mode decomposition is obtained during this post-processing procedure, the same set of images can be used to decompose a beam in different sets of spatial modes. As demonstrated here, the experimental implementation of our approach requires a simple interferometric scheme and minimal equipment. Hence, it may be readily extended to current experiments dealing with the characterization of spatial properties and OAM decomposition of structured light. 

\section*{Methods}
{\bf A. Spatial modes carrying OAM}\\[1 ex]
Using adimensional cylindrical coordinates $\rho=r/w_0$ and $\zeta=z/z_R$, where $w_0$ is the waist radius of the Gaussian envelope and $z_R$ the Rayleygh range, respectively, Laguerre-Gaussian $LG_{p,m}$ modes have the well known expression:
\begin{equation}
\begin{split}\label{eq:LGmodes}
LG_{p,m}(\rho,\zeta,\phi)=&\sqrt{\frac{2^{|m|+1} p!}{\pi (p+|m|)!(1+\zeta^2)}}\Biggr(\frac{\rho}{\sqrt{1+\zeta^2}}\Biggr)^{|m|}\times\\&e^{-\frac{\rho^2}{1+\zeta^2}}L_{p}^{|m|}(2\rho^2/(1+\zeta^2))\times\\&
e^{i\frac{\rho^2}{\zeta+1/\zeta}}e^{i m\phi-i (2p+|m|+1)\arctan(\zeta)},
\end{split}
\end{equation}
where $L_{p}^{|m|}(x)$ is the generalized Laguerre polynomial, $p$ is a positive integer and $m$ is the azimuthal index associated with the OAM.

When a Gaussian beam passes through an optical element that impinges on it a phase factor $e^{im\phi}$, the outgoing field is described by a Hypergeometric-Gaussian mode $\text{HyGG}_{p,m}$ \cite{Karimi2007,Karimi2009} with $p=-|m|$:
\begin{equation}
\begin{split}\label{eq:hgmodes}
\hg{p,m}(\rho,\zeta,\phi)=&\sqrt{\frac{2^{1+|m|+p}}{\pi\Gamma(1+|m|+p)}}\frac{\Gamma(1+|m|+p/2)}{\Gamma(1+|m|)}\times\\&
i^{|m|+1}\zeta^{p/2}(\zeta+i)^{-(1+|m|+p/2)}\times\\&\rho^{|m|}
e^{-i\rho^2/(\zeta+i)+im\phi}\times \\ & _{1}F_{1}\biggr(-p/2;|m|+1;\rho^2/(\zeta(\zeta+i))\biggr),
\end{split}
\end{equation} 
where $\Gamma(z)$ is the Euler Gamma function and $_{1}F_{1}(a;b;z)$ is the confluent Hypergeometric function.\\[1 ex]

{\bf B. Generating structured light using $q$-plates}\\[1 ex]
A $q$-plate is formed by a thin layer of liquid crystals; the angle $\alpha$ describing the orientation of the optic axis of such molecules is a linear function of the azimuthal angle, that is $\alpha(\phi)=\alpha_0+q\phi$ ($q$ is the topological charge). In our experiments we set $\alpha_0=0$. In this case, the action of the $q$-plate is described by the following Jones matrix (in the basis of circular polarizations):
\begin{equation}\begin{split}
\hat{Q}(\delta)=&\cos\biggr (\frac{\delta}{2}\biggr)\begin{pmatrix}1&0\\0&1\end{pmatrix}+\\& i \sin\biggr (\frac{\delta}{2}\biggr)\begin{pmatrix}0&e^{-i m\phi}\\ e^{im\phi}&0\end{pmatrix},
\end{split}
\label{eq: q-plate}\end{equation}
where $m=2q$ and $\delta$ is the plate optical retardation, controllable by applying an external electric field \cite{Piccirillo2010}. It is worth noting that the second term of Eq.\ \ref{eq: q-plate} introduces the azimuthal dependance associated with the OAM degree of freedom. When a left or right circularly polarized Gaussian beam passes through a $q$-plate with $\delta=\pi$, positioned at the waist of the beam, the output beam is given by HyGG$_{-|m|,\pm m}$, respectively \cite{Karimi2009}. In our experiment we generated superpositions of several OAM modes using single or cascaded $q$-plates, characterized by specific values of $q$ and $\delta$ that are reported in the figure captions.\\[1 ex]

{\bf C. Limitations on the set of detectable spatial modes.}\\[1 ex]
We briefly discussed in the main text that the finite size of the detector area and the finite dimension of sensor pixels impose certain restrictions on the features of the helical modes that can be resolved in our setup. Let us consider the simple case wherein we want to decompose the signal field in terms of LG$_{p,m}$ modes, and we want to evaluate the $p,m$-bandwidth of detectable modes. We consider only the case $m>0$, since only the absolute value $|m|$ is relevant to our discussion. Consider a camera with $N\times N$ pixels, with pixel dimensions $d\times d$ (in our setup $N=576$ and $d=9$ $\mu$m). We define the following quantities:
\begin{align}\label{eq:LG_bounds}
r_\text{max}&=d\,N/2, \\
r_\text{min}&=m\,d/\pi,\\
r_1&=w(z) \times \nonumber \\ 
&\sqrt{\frac{2p+m-2-[1+4(p-1)(p+m-1)]^{1/2}}{2}},\\
r_p&=w(z)\times \nonumber \\
&\sqrt{\frac{2p+m-2+[1+4(p-1)(p+m-1)]^{1/2}}{2}},\\
\tilde r_p&=w(z)\sqrt{2p+m+1}.
\end{align}
Here $r_\text{max}$ is the maximum radius available on the sensor; $r_\text{min}$ is the minimum radial distance where azimuthal oscillation associated with the OAM content of the LG$_{p,m}$ mode can be detected, before facing aliasing issues; $r_1$ is a lower bound for the first root of the Laguerre polynomials contained in the expression of LG modes; similarly, $r_p$ is the upper bound for the $p-$th root, while $\tilde r_p$, with $r_p<\tilde r_p$, delimits the oscillatory region of the Laguerre polynomials \cite{Ismail1992,Gatteschi2002}. Interestingly, the spatial region $r_1<r<\tilde r_p$ well approximates the area containing all the power associated with the mode. At the same time, the quantity $\Lambda=(r_p-r_1)/p$ well describes the average distance between consecutive nodes of the LG mode, defining the periodicity of their radial oscillations. A given LG$_{p,m}$ mode is then ``detectable'' (or properly ``resolvable'') if all the following conditions are satisfied:
\begin{equation}\label{eq:pm_system}
\left\{
\begin{array}{ll}
r_\text{min}<r_1&\text{(i)}\\
r_p<r_\text{max}&\text{(ii)}\\
\Lambda>2d&\text{(iii)}
\end{array}\right.
\end{equation}
Indeed, we are requiring that (i) the field is vanishing below the azimuthal aliasing threshold given by $r_\text{min}$, that (ii) all the power associated with the mode is contained in the sensor area and (iii) that the field radial oscillations have a spatial period such that at least two pixels are contained in a single period, respectively (radial aliasing limit). It is easy to check that in our configuration, where the beam waist is $w(z)=0.4$ mm, conditions (i) and (iii) are always satisfied for the values of $\{p,m\}$ that are solution of (ii), i.e. the limiting factor is only the dimension of the sensor area. By solving such inequality, we get the relation
\begin{align}
p<\left(\frac{N^2d^2}{4\,w^2(z)}-m-1\right)/2
\end{align} 
In Fig.\ \ref{fig:LG_detectable} we plot a colormap for a rapid visualization of detectable modes. If we apply this analysis to the case of Fig.\ \ref{fig:LGexpansion}, in which a beam with $m=8$ is studied, we obtain that only radial modes with $p<16$ can be detected. In general, for smaller values of $w(z)$ the determination of detectable LG mode is more complex and requires the complete resolution of the system of inequalities system given in (\ref{eq:pm_system}).

\section*{Funding Information}
This work was supported by the European Research Council (ERC), under grant no.694683 (PHOSPhOR).




%

 \end{document}